\documentclass[aps,reprint,prl,superscriptaddress,floats,floatfix]{revtex4-2}

\usepackage{amsmath}
\usepackage{amsfonts}
\usepackage{amssymb}
\usepackage{graphicx}
\usepackage{color}
\usepackage{tikz}
\usepackage[colorlinks=true,citecolor=blue,linkcolor=blue,urlcolor=blue]{hyperref}
\usepackage{times}
\usepackage{amstext}
\usepackage{latexsym}
\usepackage[running,mathlines]{lineno}
\usepackage{bbm}
\usepackage{verbatim}

\providecommand{\bra}[1]{\langle #1 \rvert}
\providecommand{\ket}[1]{\lvert #1 \rangle}

\begin{document}
\raggedbottom

\title{Bright and dark states of light: The quantum origin of classical interference}

\author{Celso J. Villas-Boas}
\affiliation{Departamento de F\'{i}sica, Universidade Federal de S\~{a}o Carlos, 13565-905 S\~{a}o Carlos, S\~{a}o Paulo, Brazil}

\author{Carlos E. M\'{a}ximo}
\affiliation{Departamento de F\'{i}sica, Universidade Federal de S\~{a}o Carlos, 13565-905 S\~{a}o Carlos, S\~{a}o Paulo, Brazil}
\affiliation{Department of Physics, ETH Zurich, 8093 Zurich, Switzerland}

\author{Paulo J. Paulino}
\affiliation{Departamento de F\'{i}sica, Universidade Federal de S\~{a}o Carlos, 13565-905 S\~{a}o Carlos, S\~{a}o Paulo, Brazil}
\affiliation{Institut für Theoretische Physik, Eberhard Karls Universität Tübingen, Auf der Morgenstelle 14, 72076 Tübingen, Germany}

\author{Romain P. Bachelard}
\affiliation{Departamento de F\'{i}sica, Universidade Federal de S\~{a}o Carlos, 13565-905 S\~{a}o Carlos, S\~{a}o Paulo, Brazil}

\author{Gerhard Rempe}
\affiliation{Max-Planck-Institut f\"ur Quantenoptik, Hans-Kopfermann-Str.\ 1, D-85748 Garching, Germany}

\begin{abstract}
\noindent Classical theory asserts that several electromagnetic waves cannot interact with matter if they interfere destructively to zero, whereas quantum mechanics predicts a nontrivial light-matter dynamics even when the average electric field vanishes. Here, we show that in quantum optics, classical interference emerges from collective bright and dark states of light, \textit{i.e.}, particular cases of two-mode binomial states, which are entangled superpositions of multi-mode photon-number states. This makes it possible to explain wave interference using the particle description of light and the superposition principle for linear systems only. It also sheds new light on an old debate concerning the origin of complementarity.
\end{abstract}

\maketitle

\noindent The quest for understanding what light is and what its properties are comes from the ancient Greek school of philosophy~\cite{darrigol2012history}. Since then, the subject has been extensively studied, with prominent contributions from Newton and Huygens, the former defending the corpuscular and the latter advocating the wave nature of light ~\cite{newton1952opticks,huygens1912treatise}. This dispute remained unresolved until, among others, Young performed experiments on light diffraction ~\cite{young1855miscellaneous, rubinowicz1957thomas} and Maxwell developed a unified theory of electromagnetism which includes a wave equation for the light field
~\cite{jackson1999classical}. This effectively removed any doubts about the wave nature of light. However, in 1905, Einstein explained the photoelectric effect by reintroducing the idea of light particles~\cite{einstein1965concerning} (later reconsidered by Lamb and Scully~\footnote{In 1968, Lamb and Scully provided a semiclassical model, which treats the electromagnetic field classically and the matter quantum mechanically, to explain the photoelectric effect without the need for the corpuscular aspect of light~\cite{lamb1968}.}). Since then, and with the advent of quantum physics, light is associated with both properties, wave and particle. Depending on the experiment, one or the other aspect manifests itself: the interference of delocalized waves or the propagation of particles along well-defined trajectories. Although this is textbook knowledge by now, it was highly debated at its time. For example, Millikan argued that the particle aspect \textit{``flies in the face of the thoroughly established facts of interference''}~\cite{Millikan1916}.

Here we resolve Millikan's objection and show that the interference between independent radiation modes, usually taken as an undoubted signature of the wave character, has a purely corpuscular explanation. For this we incorporate a full quantum-mechanical description of the measurement process in terms of energy exchange between the light and a sensor atom, 
and carry out a detailed analysis in terms of multi-mode collective states of the radiation field. This allows us to identify states that can be coupled with matter, dubbed bright states, and states that do not couple with matter, called dark states. Such states are also known as sub- or super-radiant states~\cite{Delanty2011}, or generalized ground states (for the dark states)~\cite{Alsing1992}, and are particular cases of two-mode binomial states~\cite{stoler1985binomial, dattoli1987binomial, gerry1991interaction, barnett1998negative, Bjork2001, bjork2002applications, gerry2005dark, stohr2023quantum, Yurke1986}. Surprisingly, such physical interpretation in terms of particle states (that either couple or do not couple with detectors) have never been used so far to explain the interference patterns that emerge, e.g., in optical double-slit experiments~\footnote{The superposition principle applies to any linear system ranging from algebraic equations and linear differential equations such as Schrödinger, Heisenberg, and wave equations to systems of equations of these forms (e.g., equations of motion of coupled harmonic oscillators). Therefore, the superposition principle is not a signature of the wave character of the radiation.}.

We then discuss the rather counter-intuitive result that a vanishing photon-detection probability at locations of destructive interference does not prove the absence of photons. We argue, instead, that these photons are in a state that is perfectly dark for the employed sensor atom. In other words, we replace Glauber's explanation of constructive or destructive interference in terms of superposed transition amplitudes~\cite{glauber2006} by a description where the light is in a state that can or cannot excite the atom, respectively. This leads to a new view on which-path detection in double-slit experiments~\cite{kwiat1992observation, walborn2002double, luo2024young} as, contrary to the standard notion, photons always reach the dark regions, independently of the presence of the detector. 

\begin{figure*}
\begin{center}
\includegraphics[width=\linewidth,height=\textheight,keepaspectratio]{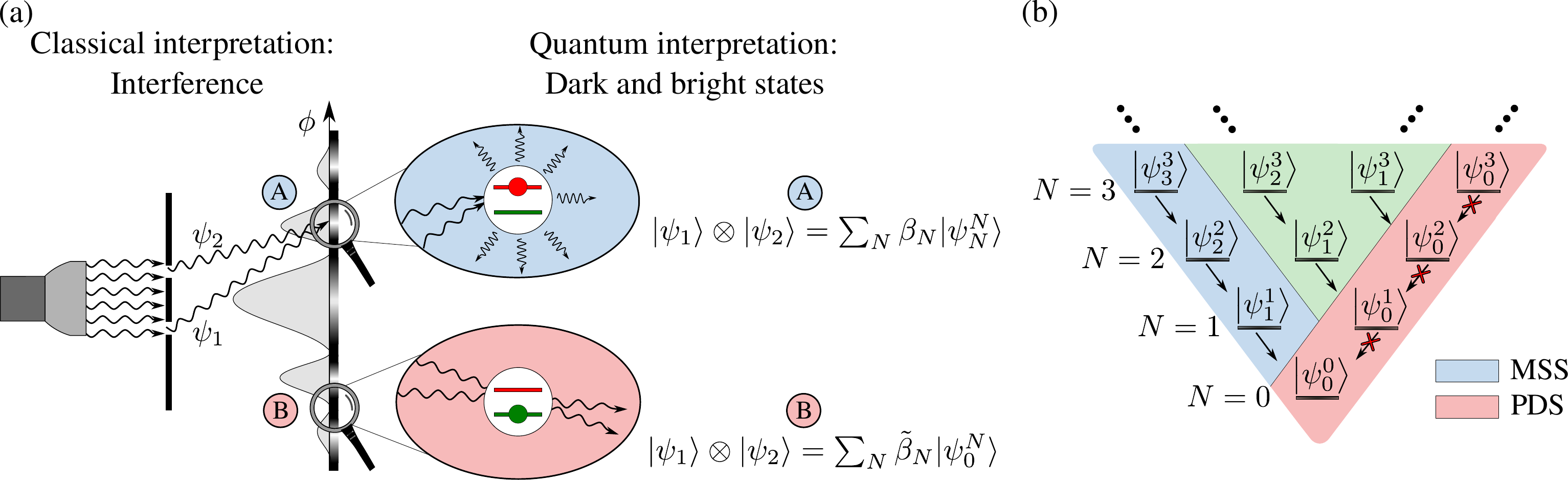}
\caption{(a) Double slit experiment: Two waves $\psi_1(\phi_1=\textbf{k}_1\cdot\textbf{r}_1)$ and $\psi_2(\phi_2=\textbf{k}_2\cdot\textbf{r}_2)$ meet to form fringes which can be interpreted in terms of interference in the classical theory, and of dark and bright states in our approach. In the latter, there is light (photons) at all points, and the dark and bright regions are explained  microscopically in terms of the interaction between the atoms and the collective states that excite (bright regions) and that do not excite (dark regions). (b) Dicke-like ladder for two light modes with the left side containing the maximally superradiant states (MSS, in blue), and the right one with the perfectly dark states (PDS, in red). Unlike the atomic case, the number of excitations is not bounded ($N\rightarrow\infty$). Also, the interaction between the collective modes of the field and matter occurs in a way that induces decays from the leftmost (bright) states to the rightmost (dark) states, so that, upon reaching the perfectly dark states, the field decays stops happening and, consequently, the field stops exciting the matter.} 

\label{fig:ladder}
\end{center}
\end{figure*}

We also show that the collective states of the light fall into three distinct classes: perfectly dark, maximally superradiant, and intermediate. Interestingly, classical interference, fully destructive or constructive, is described by a superposition of perfectly dark or maximally superradiant states only (see the illustration in Fig.\ref{fig:ladder}(a)), but any intermediate quantum state has no counterpart in classical theory, a feature which could be verified in measurements of the first-order correlation function only \cite{Steuernagel2001}.

According to the quantum theory of optical coherence introduced by Glauber~\cite{Glauber1963_1, glauber2007quantum}, the statistical properties of a given field can be derived from its electrical field operator
\begin{equation}
    \textbf{E}(\textbf{r},t) = \textbf{E}^{(+)}(\textbf{r},t)+\textbf{E}^{(-)}(\textbf{r},t),
\end{equation}
with $\textbf{E}^{(+)}(\textbf{r},t)$ and $\textbf{E}^{(-)}(\textbf{r},t)$ the positive and negative frequency parts, respectively. Quantum mechanically, these parts are proportional to the photon annihilation and creation operators, that is, $\textbf{E}^{(+)}(\textbf{r},t) \propto a$ and $\textbf{E}^{(-)}(\textbf{r},t) \propto a^{\dagger}$. Still according to~\cite{Glauber1963_1, glauber2007quantum}, the probability of a photon from a single mode in a given state $\ket{\Psi}$ being absorbed by a sensor atom is proportional to 
\begin{equation}
\bra{\Psi}\textbf{E}^{(-)}\textbf{E}^{(+)}\ket{\Psi}\propto\bra{\Psi}a^{\dagger}a\ket{\Psi}.
\label{prob_exc}
\end{equation}
This expression comes from the energy-exchange interaction between the field and the sensor, described by the Hamiltonian (in the rotating-wave approximation) 
\begin{equation}
H= \textbf{E}^{(+)}(\textbf{r},t)\sigma^+ + \textbf{E}^{(-)}(\textbf{r},t)\sigma^-,
\label{H0}
\end{equation} 
with $\sigma^+$ ($\sigma^-$) the raising (lowering) operators that induce transitions between ground $\ket{g}$ and excited $\ket{e}$ states of the sensor atom. From Eq. (\ref{prob_exc}) one can easily see that, for single mode fields, only  zero-intensity (vacuum) fields are unable to excite the sensor. 

For two-mode fields, and from a quantum perspective, the situation is much more interesting. To see this, 
we evaluate the probability of exciting the sensor 
using the eigenstates of the positive-frequency operator with null eigenvalues, that is, $\textbf{E}^{(+)}(\textbf{r},t)\ket{\Psi}=0$ or, equivalently, $H\ket{\Psi}\ket{g} = 0$. For a single mode, only the vacuum state satisfies this condition. But as shown below, multi-mode fields allow for a plethora of such states, even states with many photons in each mode. In the context of cavity quantum electrodynamics, the states which carry photons but are unable to excite an atom were dubbed ``generalized ground states''~\cite{Alsing1992}, but here we decide to name them perfectly dark states (PDSs) since the sensor cannot see the field whenever it is in such a state.

To address the multi-mode case, we consider two modes, $A$ and $B$, represented by their respective annihilation (creation) operators $a$ and $b$ ($a^{\dagger}$ and $b^{\dagger}$), and a relative phase between them given by $\theta$. In this case, the positive frequency operator can be written as 
\begin{equation}
\textbf{E}^{(+)}(\textbf{r},t) \propto (a+be^{i\theta})    
\end{equation}
and, consequently, the Hamiltonian (\ref{H0}) becomes $ H= g(a+be^{i\theta})\sigma^{+}+g(a^\dagger+b^\dagger e^{-i\theta})\sigma^-$. Here, $g= g(\textbf{r})$ denotes the coupling constant between the sensor and the two modes, assumed equal for both modes for simplicity, and $\textbf{r}$ is the sensor (atom) position. Then, we introduce the symmetric and antisymmetric collective operators $c = (a+b e^{i\theta})/\sqrt{2}$ and $d = (-a e^{-i\theta}+b)/\sqrt{2}$~\cite{bjork2002applications, moussa2014}, respectively. For any number $n \le N$, we then write the states with a total number of $N$ excitations $\ket{\psi^N_n (\theta)}$ in the collective ($\{c,d\}$) and in the bare ($\{a,b\}$) basis as {\cite{dattoli1987binomial, Bjork2001, bjork2002applications} 
\begin{flalign}
|\psi_{n}^{N}(\theta)\rangle  & \equiv\left|n,N-n\right\rangle _{c,d} =\frac{\left(c^{\dagger}\right)^{n}\left(d^{\dagger}\right)^{N-n}}{\sqrt{n!\left(N-n\right)!}}\left|0,0\right\rangle_{c,d}\nonumber 
\\
 & =\frac{\left(a^{\dagger}+b^{\dagger} e^{-i\theta}
 \right)^{n}\left(-a^{\dagger} e^{i\theta}+b^{\dagger}\right)^{N-n}}{\sqrt{2^{N}n!\left(N-n\right)!}}\left|0,0\right\rangle_{a,b}.
 \label{psigeral}
\end{flalign}

As shown in the supplemental material (SM), which includes the Refs.~\cite{scully1997, mandel1995,Zavatta2004, Parigi2007,Cohen1992,Murr2006}, the states $\ket{\psi^N_n}$ constitute a complete basis satisfying $H \left|\psi_{n}^{N}\right\rangle \ket{g}= g\sqrt{2n} \left|\psi_{n-1}^{N-1}\right\rangle \ket{e}$. This describes the excitation of the atom accompanied by a transition from state $\ket{\psi_{n}^{N}}$ to $\ket{\psi_{n-1}^{N-1}}$, as illustrated in Fig. \ref{fig:ladder} (b), analog to the $N$-excitation Dicke basis for multi-atom systems~\cite{Dicke1954,Delanty2011}. In analogy to the ``cooperation number'' for Dicke states, the $\sqrt{2n}$ factor represents the cooperativity of the absorption, where the $\sqrt{2}$ appears due to the two modes. 

Of particular interest is the case $n=0$ for which $\textbf{E}^{(+)}(\textbf{r},t)\ket{\psi_0^{N}}=0$. The $n=0$ state is therefore unable to excite the sensor atom, for any $N$, and this is why we name it the perfectly dark state (PDS) for the subspace of $N$ photons \cite{bjork2002applications, Delanty2011}
\begin{equation}
\left|\psi_0^{N} (\theta)\right\rangle =\sqrt{\frac{N!}{2^{N}}}\sum_{m=0}^{N}\frac{(-1)^{m}e^{im\theta}}{\sqrt{m!\left(N-m\right)!}}\left\vert m,N-m\right\rangle_{a,b}. \label{dark}
\end{equation}
In atomic system it was coined subradiant by Dicke since $H \ket{\psi_{0}^{N}} \ket{g} =0$. Oppositely, the $n=N$ state {\cite{bjork2002applications, Delanty2011} 
\begin{equation}
\small{
\left|\psi_N^{N} (\theta)\right\rangle =e^{-iN\theta}\sqrt{\frac{N!}{2^{N}}}\sum_{m=0}^{N}\frac{e^{im\theta}}{\sqrt{m!\left(N-m\right)!}}\left\vert m,N-m\right\rangle_{a,b}
}
\label{sup}
\end{equation}
comes with a transition rate $g\sqrt{2N}$, which is $\sqrt{2}$ times that of the single-mode result: $H_{JC}\ket{g}\ket{N}=g\sqrt{N}\ket{e}\ket{N-1}$, with $H_{JC}$ denoting the standard Jaynes-Cummings Hamiltonian~\cite{Jaynes1963}. State \eqref{sup} is the analogue of the symmetric superradiant mode, studied by Dicke in the atomic decay cascade~\cite{Dicke1954,Gross1982}, and represents the most superradiant of the states with $N$ photons \cite{Delanty2011}. We here refer to this state as a maximally superradiant state (MSS) or bright state. Finally, states within the range $0<n<N$ have intermediate transition rates. In contrast to the two-level-atom case, the present Hilbert space is unbounded, even for a finite number of field modes, since each one can support an arbitrary number of photons~\cite{Delanty2011} (see Fig.\ref{fig:ladder}(b)). 

As we have just seen, dark states are eigenstates of the operator $\textbf{E}^{(+)}(\textbf{r},t)$ with null eigenvalue, which implies that they are undetectable by usual sensors such as those described by two-level atoms. On the other hand, the bright states couple stronger than in the single-mode case. Thus, a natural question arises concerning the connection between the quantum mechanical dark (bright) states and the classical effects of destructive (constructive) interference between radiation fields, as in regions of destructive interference, no light is detected, while in regions of constructive interference, light scattering is enhanced. To address this question we consider, without loss of generality, the case of two modes with $\theta =0$. Then, 
one can easily show that in-phase coherent states decompose exclusively on the MSS subspace: 
\begin{equation}
\ket{\alpha,\alpha}=e^{-|\alpha|^{2}}\sum_{N=0}^{\infty}\sqrt{\frac{2^{N}}{N!}}\alpha^{N} \ket{\psi_N^N}.\label{eq:coh1}
\end{equation}
This implies an enhanced absorption by a factor of $2$ ($H\ket{\alpha,\alpha}\ket{g}=2\alpha g\ket{\alpha,\alpha}\ket{e}$) as compared to a single coherent state ($H\ket{\alpha}\ket{g}=g\alpha\ket{\alpha}\ket{e}$). The two-mode MSS quantum state, therefore, corresponds to the constructive interference of classical in-phase fields. 

On the other hand, two coherent fields with opposite phases decompose in terms of PDSs only:
\begin{equation}
\ket{\alpha,-\alpha} =e^{-|\alpha|^{2}}\sum_{N=0}^{\infty}\sqrt{\frac{2^{N}}{N!}}\alpha^{N}\ket{\psi_0^N}.\label{eq:coh2}
\end{equation}
This state gives a suppressed interaction $H\ket{\alpha,-\alpha}\ket{g}=0$, which can be interpreted either as belonging to the PDS subspace or, classically, as a destructive interference for two fields with the same amplitude  but opposite phases. However, not every destructively interfering field is undetectable. This can be seen by considering two modes in the state 
\begin{equation}\ket{\Upsilon}=\left(\ket{0}_a+\ket{1}_a\right)\left(\ket{0}_b-\ket{1}_b\right)/2.
\end{equation}
In each of the modes, we have a non-zero average electric field, but as they are out of phase with each other, the average resulting electric field is zero. According to the classical interpretation of interference, such a field would be undetectable but, according to Glauber's theory~\cite{Glauber1963}, it is detectable, in the sense that it will induce a non-trivial dynamics for the sensor/atom. This can be easily explained using the description in terms of bright and dark states, since such state can be written in the form $\ket{\Upsilon} = \left[ \ket{\psi_0^0} - \sqrt{2}\ket{\psi_0^1} +  \left(\ket{ \psi_0^2} - \ket{\psi_2^2} \right)/\sqrt{2} \right]/2$, which shows a projection onto the detectable subspace of bright states.

A key phenomenon for evidencing the wave nature of light comes from Young's double-slit experiment. The fundamental result is that both classical and single-photon coherent sources produce the same fringe pattern, despite the very different nature of these fields \cite{feynman1965feynman, kocsis2011observing, aspden2016video, luo2024young}. To revisit this experiment using the collective basis, one can consider two equally weighted light modes emerging from two slits and in the far-field limit. Without loss of generality, we assume both waves with wave vectors $\textbf{k}_1$ (mode $a$) and $\textbf{k}_2$ (mode $b$), with $|\textbf{k}_1|=|\textbf{k}_2|=k$. Then, $\textbf{k}_1\cdot\textbf{r}_1$ and $\textbf{k}_2\cdot\textbf{r}_2$, with $\textbf{r}_i$ the vector connecting the $i^\text{th}$ slit with the sensor position, are the phases acquired by the respective fields when propagating from slits 1 and 2, respectively, to the detection point (see Fig.~\ref{fig:ladder}).

For a single photon impinging on a double slit, the field in the plane of interest and the detection process can be described by replacing the two slits with two source atoms, the first at the position $\textbf{d}_1$ and the second at the position $\textbf{d}_2$ (the positions of the slits)~\cite{scully1997}. Apart from a normalization factor which depends on the radiation pattern of the two `slit' atoms, $g(\textbf{k})$, the field is given by the state (see SM for details)~\textcolor{red}{\cite{Walls1977,scully1997}}

\begin{equation}
\ket{S} = \frac{1}{\sqrt{2}}\left( 
e^{-i\textbf{k}_1 \cdot \textbf{d}_1 }\ket{1,0}_{a,b} + e^{-i\textbf{k}_2 \cdot \textbf{d}_2 }\ket{0,1}_{a,b}
\right) \small{\bigotimes_{\textbf{k}\ne \textbf{k}_1,\textbf{k}_2 }} \ket{0_{\textbf{k}}}.
\label{psidoubleslit}
\end{equation}

In previous works~\cite{scully1997,Walls1977}, all the discussions on the interference pattern are restricted to the probability of detecting light at position $\textbf{r}$, which is given by the first-order intensity correlation function
%
$G^{(1)}(\textbf{r},\textbf{r},0) = \bra{\psi}E^{(-)}E^{(+)}\ket{\psi}= \left|\bra{0}E^{(+)}\ket{\psi} \right|^2$,
with $E^{(+)} \propto a e^{i\textbf{k}_1 \cdot \textbf{r}} + b e^{i\textbf{k}_2 \cdot \textbf{r}} \propto a + b e^{i\theta}$, and $\theta = (\textbf{k}_2-\textbf{k}_1)\cdot \textbf{r}$, without engaging in any discussion regarding the physical meaning of the collective states at the detector's position—specifically, without addressing the existence of states that couple to the detectors and others that do not, as is done here.} As described above, for this collective measurement operator the dark and bright states are 
$\ket{\psi^1_0(\theta)} = \left(\ket{1,0}_
{a,b} - e^{-i\theta }\ket{0,1}_{a,b}
\right)/\sqrt{2}$ and 
$\ket{\psi^1_1(\theta)} = \left(\ket{1,0}_
{a,b} + e^{-i\theta }\ket{0,1}_{a,b}
\right)/\sqrt{2}$, respectively (apart from the other vacuum modes of the electromagnetic field). 
%
With such equations, we can rewrite Eq. (\ref{psidoubleslit}), up to a global phase factor as
\begin{eqnarray}
\small{
\ket{S'}= 
\text{cos}\left(\delta\phi/2\right)\ket{\psi^1_1(\theta)} - i \text{sin}\left(\delta \phi/2 \right) \ket{\psi^1_0(\theta)}
},
\label{Sfinal}
\end{eqnarray}
where $\delta\phi= -( \textbf{k}_2 \cdot \textbf{d}_2 - \textbf{k}_1 \cdot \textbf{d}_1) +\theta= \textbf{k}_2 \cdot \textbf{r}_2 - \textbf{k}_1 \cdot \textbf{r}_1 $,  
\textit{i.e.}, $\delta\phi$ represents the phase difference of the two light paths from the slits to the sensor atom. Clearly, at any detector position ($\delta\phi$) we may have a bright, a dark, or a superposition of bright and dark states, which implies that the photon can be at any position. In other words, apart from the $g(\textbf{k})$ distribution, the average number of photons as a function of $\delta \phi$ is constant, \textit{i.e.}, $\langle a^{\dagger}a + b^{\dagger}b \rangle (\delta \phi) = 1$. However, the sensor atom can detect the photon only at positions where the bright state survives. 

Our finding is consistent with the controversially discussed~\cite{Scully1991,Storey1994}, but experimentally confirmed~\cite{Durr1998}, observation that the interference pattern, which is the momentum distribution in the far field of the double slit, can be changed by means of a which-path detector that is subtle enough not to impart momentum kicks on the photon. Since, in our description, the photon can, in principle, arrive at any position on the screen, washing out the interference fringes does not require the which-path detector to steer the photon from a bright into a dark region of the interference pattern. Instead, it is sufficient for the detector to destroy the purity of the dark state. This makes the photon detectable in the otherwise dark region. The results reported here are therefore further confirmation of the view that quantum-mechanical complementarity is based on the abstract concept of entanglement instead of the intuitive notion of random momentum kicks induced by which-path detectors~\cite{Scully1991}.

On the other hand, when a coherent state is sent to a double slit, part of the light goes through slit 1 at position $\textbf{d}_1$, and part of it goes through slit 2 at position $\textbf{d}_2$. In this case, we do not have a superposition state, but rather a product state of the two modes $\textbf{k}_1$ and $\textbf{k}_2$ in coherent states with amplitude $\alpha$ (assumed the same for both slits) and a relative phase which depends on the position of the slit, \textit{i.e.}
\begin{equation}
\ket{e^{-i\textbf{k}_1\cdot\textbf{d}_1} \alpha, e^{-i\textbf{k}_2\cdot\textbf{d}_2}\alpha}_{a,b} =  e^{-|\alpha|^{2}}\sum_{N=0}^{\infty}C_N \ket{\chi^{N}(\delta\phi)},
\label{coh}
\end{equation}
with $C_N = \sqrt{2^N/N!}(e^{-i\textbf{k}_2\textbf{d}_{2}}\alpha)^{N}$ and the phase-dependent state
\begin{equation}
 \left|\chi^{N}\left(\delta\phi\right)\right\rangle =\sqrt{\frac{N!}{2^{N}}}\sum_{m=0}^{N}\frac{e^{-i m\delta\phi}e^{i m\theta}}{\sqrt{m!\left(N-m\right)!}}\left\vert m,N-m\right\rangle,
\end{equation}
with $\delta \phi$ defined right after Eq. (\ref{Sfinal}). Such state corresponds to a MSS when the two modes are in phase, $\left|\chi^{N}\left(2l\pi\right)\right\rangle =e^{iN\theta}\ket{\psi_{N}^{N}(\theta)}$, and to a PDS when in opposite phases, $\left|\chi^{N}\left((2l+1)\pi\right)\right\rangle =\ket{\psi_{0}^{N}(\theta)}$, with $l = 0,\pm 1, \pm 2,...$\,. In this case, the average number of photons is again independent of the phase difference, \textit{i.e.}, 
\begin{equation}
\langle a^{\dagger}a + b^{\dagger}b \rangle (\delta \phi) = 2|\alpha|^2, 
\end{equation}
meaning that photons are present at every point on the screen, contrary to the standard classical description of interference, which states that no light arrives at points of destructive interference. The single-photon state $\ensuremath{\left|S\right\rangle}$, which decomposes as a sum of PDSs or MSSs only (see Eq.~\eqref{Sfinal}), has the same feature, as discussed above. Therefore, the decomposition either in only PDSs or only MSSs explains why single-photon fields and classical fields exhibit the same fringe pattern. 

The case discussed highlights a general result: as shown in the SM, states of light composed solely of PDSs or MSSs produce the same interference patterns as those from linear (classical) optics or coherent states. However, this does not hold for general collective states $\ket{\psi^N_n(\theta)}$ (Eq. \ref{psigeral}) or superpositions containing both dark and bright states. This distinction can help differentiate quantum from classical states of light without relying on field-field correlations~\cite{mandel1995, scully1997}. For example, consider a Mach-Zehnder interferometer (MZI) with an input state $\ket{\psi^2_1}=(\ket{0,2}{a,b} - \ket{2,0}{a,b})/\sqrt{2}$, generated by sending two photons into a $50/50$ beam splitter~\cite{Hong1987}. In this setup, the two output ports of the MZI contain the same average photon number, independent of the phase $\varphi$, with $\langle n_{a} \rangle = \langle n_{b} \rangle = 1$. This eliminates interference fringes, reducing visibility to zero, unlike classical fields, where intensity in one output port can vanish depending on $\varphi$. This reduced visibility in the quantum case reflects the quantum nature of the field.

In light of our findings, we emphasize that the single-mode case has a unique PDS, namely the vacuum state $\ket{0}$. It may sound trivial since there is no photon to excite the detector, yet its interest lies in its uniqueness: any other state excites the sensor atom. Thus, our interpretation in terms of dark and bright states provides a new way to explain why single-mode Fock states $\ket{N}$ with $N>1$ do excite the sensor atom, even for zero mean electric fields. The multi-mode case, however, is fundamentally different since it possesses an infinite family of dark states with an arbitrarily large number of photons, which do not couple to the sensor atom in the ground state. In addition, the two-mode case also predicts bright and intermediate states, the latter having no correspondence in classical physics. The above discussion, originally for two radiation field modes, extends directly to $M$ modes/slits~\cite{sinha2010ruling, magana2016exotic} (see SM), where any interference is described via collective bright, dark, and intermediate states. A pulsed light from mode-locked lasers exemplifies this, with photons forming bright states during pulses and dark states between them~\cite{diniz2024pulsed}. 

From an experimental perspective, the two-mode light-matter interaction discussed here suggests an implementation in optical cavities coupled to a two-level atom~\cite{Hamsen2018, Brekenfeld2020}, trapped ions where a single emitter can be coupled to its two vibrational modes~\cite{Wineland2003,Markus2020} as performed in \cite{parke2024phononic}, or superconducting circuits~\cite{Wallraff2021}. We visualize many possibilities, drawing inspiration from the diverse applications that appear in the context of super- and subradiance in atomic systems. For example, one could employ photonic superradiant states to further enhance light emission in high-brightness light sources. On the other hand, as the dark states do not interact with matter, they could, in principle, be employed as decoherence-free photonic quantum memories. Finally, by taking advantage of the fact that collective bright (dark) states do (not) interact with atoms, one could use such states to imprint a conditional phase on an atomic system, thus implementing single-shot logic operations in crossed-cavity setups~\cite{Brekenfeld2020}, 
allowing for universal quantum computing with traveling modes~\cite{Solak2024}.

In conclusion, we have discussed how a description of multi-mode light in terms of maximally superradiant or perfectly dark collective states offers a natural interpretation for constructive and destructive interference. Remarkably, this Dicke-like bosonic basis applies to classical and non-classical states of light, thus going beyond the simple classical approach of average fields. We have shown that, from a quantum perspective, interference is intimately related to the coupling of light and matter which differs for the bright and dark states. This is completely different from the classical description, where no assumption on the matter is necessary to describe the sum of electromagnetic fields. One can interpret this as a manifestation of the quantum-measurement process where the expectation value of an observable depends on the measuring apparatus~\cite{davies1981,dodonov2005}. Within this framework, we have interpreted the double-slit experiment and the interference of light waves in general in terms of bright and dark states, \textit{i.e.}, using only the corpuscular description of the light and the quantum-mechanical superposition principle. 

\begin{acknowledgments}
C.J.V.-B., C.E.M, and R.B., and are supported by the S\~ao Paulo Research Foundation (FAPESP, Grants Nos. 2017/13250-6, 2018/15554-5, 2019/11999-5, 2018/22402-7, 2023/01213-0, 2023/03300-7, and 2022/00209-6) and by the Brazilian National Council for Scientific and Technological Development (CNPq, Grants Nos. 201765/2020-9, 402660/2019-6, 313886/2020-2, 409946/2018-4, 307077/2018-7, 465469/2014-0, and 311612/2021-0). C.E.M. acknowledges funding from the SNF, project number IZBRZ2-186312/1. We thank Tobias Donner and Markus Hennrich for their comments and Wolfgang Schleich for discussions concerning the formal description of the double-slit experiment. C.J.V.-B. and R.B. thank the Max Planck Institute of Quantum Optics for their hospitality.
\end{acknowledgments}

\bibliographystyle{apsrev4-2}
\bibliography{references.bib}

\end{document}